\documentclass[sn-mathphys-num]{sn-jnl}

\usepackage{graphicx}%
\usepackage{multirow}%
\usepackage{amsmath,amssymb,amsfonts}%
\usepackage{amsthm}%
\usepackage{mathrsfs}%
\usepackage[title]{appendix}%
\usepackage{xcolor}%
\usepackage{textcomp}%
\usepackage{manyfoot}%
\usepackage{booktabs}%
\usepackage{algorithm}%
\usepackage{algorithmicx}%
\usepackage{algpseudocode}%
\usepackage{listings}%
\usepackage{physics}
\usepackage{lipsum}
\usepackage{accents}

\newcommand{\ut}[1]{\underaccent{\tilde}{#1}}
\renewcommand{\vec}[1]{\ut{#1}}

\theoremstyle{thmstyleone}%
\theoremstyle{thmstyletwo}%

\theoremstyle{thmstylethree}%

\begin{document}

\title[Article Title]{ Generalized Quantum Hadamard Test  for Machine Learning}


\author[1]{\fnm{Vivek} \sur{Mehta}}\email{vivek\_1921ph05@iitp.ac.in}
\author[1]{\fnm{Arghya} \sur{Choudhury}}\email{arghya@iitp.ac.in}
\author*[1]{\fnm{Utpal} \sur{Roy}}\email{uroy@iitp.ac.in}


\affil*[1]{\orgdiv{Department of Physics}, \orgname{Indian Institute of Technology Patna}, \orgaddress{\street{Bihta}, \city{Patna}, \postcode{801106}, \state{Bihar}, \country{India}}}




\abstract{  Quantum machine learning models are designed for performing learning tasks.  Some quantum classifier models are proposed to assign classes of inputs based on fidelity measurements. Quantum Hadamard test is a well-known quantum algorithm for computing these fidelities. However, the basic requirement for deploying the quantum Hadamard test maps input space to $L2$-normalize vector space. Consequently, computed fidelities correspond to cosine similarities in mapped input space. We propose a quantum Hadamard test with the additional capability to compute the inner product in bounded input space, which refers to the Generalized Quantum Hadamard test. It incorporates not only $ L2$-normalization of input space but also other standardization methods, such as Min-max normalization. This capability is raised due to different quantum feature mapping and unitary evolution of the mapped quantum state. We discuss the quantum circuital implementation of our algorithm and establish this circuit design through numerical simulation. Our circuital architecture is efficient in terms of computational complexities. We show the application of our algorithm by integrating it with two classical machine learning models: Logistic regression binary classifier and Centroid-based binary classifier and solve four classification problems over two public-benchmark datasets and two artificial datasets.}


\maketitle

\section{Introduction}\label{sec1}
Quantum machine learning emerges as a field where we exploit the mysterious phenomena of quantum mechanics, specifically quantum superposition and entanglement, to provide quantum advantages for solving machine learning problems with less computational complexity than classical machine learning models \cite{biamonte2017quantum,schuld2021machine}. Hence, we see that quantum mechanics is applied other than its conventional purpose for describing the microscopic physical system.  A variety of quantum classifiers are tested in quantum machine learning such as quantum algebraic-based classifiers 
 \cite{rebentrost2014quantum,lloyd2014quantum,li2015experimental,schuld2016prediction}, quantum centroid-based classifiers \cite{schuld2017implementing, blank2022compact,das2023quantum}, quantum neural networks \cite{schuld2015simulating,mangini2020quantum,rebentrost2018quantum,zhao2019building,shao2020data,yan2020nonlinear}, and variational quantum classifiers\cite{benedetti2019parameterized, mitarai2018quantum,havlivcek2019supervised,caro2022generalization,huang2021variational}. Often, variational classifiers are designated as quantum neural network models.  In particular, in quantum centroid-based classifiers \cite{park2020theory, blank2020quantum}, we measure the similarities of unseen input to each input of both classes and assign the label to that class whose inputs have high similarity with it. These similarities are usually determined by cosine similarity since we use \emph{amplitude encoding} to map the input space to the quantum Hilbert space.  $L2$-normalization of input space is the classical pre-processing requirement for amplitude encoding. Once the classical information is embedded into the probability amplitude of the basis of quantum states, we compute their similarities via fidelities using the quantum algorithm, known as the quantum Hadamard test. This amplitude encoding is used not only with quantum centroid-based classifiers but also with other classes of quantum classifiers too \cite{schuld2020circuit}.

In this paper, we propose the quantum Hadamard test with additional capability for machine learning that computes the inner product in the bounded real input space where cosine similarity is taken as a \emph{special case} when our input space is $L2$ normalized \cite{park2020theory,rethinasamy2023estimating}. Bounded real input space means components of any vector that belongs to this space should be bounded between -1 to 1. This version of the quantum Hadamard test refers to the Generalized quantum Hadamard test (GQHT) since it generalizes the utility of the already well-known quantum Hadamard test (QHT) beyond $L2$ normalized inputs. For instance, we can state that GQHT incorporates not only $L2$ normalization of input space but also other standardization techniques, such as Min-max normalization for machine learning tasks,  since in some cases $L2$ normalization of inputs which is a basic requirement of QHT, changes the distribution of inputs in the input space, consequently important information may also be destroyed due to this normalization, while it does not happen with the Min-max normalization method. We validate the effectiveness of our GQHT algorithm for classification tasks over the public-benchmark datasets (Iris and Seeds) as well as synthesized datasets (Blobs and Two Half Moons), by incorporating it as subordinate with two linear classifiers:  a parametric classification algorithm, \textit{Logistic regression binary classifier} \cite{bishop2023deep} and a non-parametric classification algorithm, \textit{Centroid-based binary classifier} \cite{scholkopf2018learning}. Our sole purpose for using these two algorithms is to show the workings of our scheme for machine learning when we consider the bounded input vector other than the $L2$ normalized ones. Beyond machine learning, we can use this GQHT algorithm anywhere, when we require an inner product between bounded vectors. The additional capability of our algorithm arises due to the exploitation of another quantum feature mapping scheme instead of the amplitude encoding scheme. We also discuss the quantum circuit architecture for implementing our encoding scheme that uses the quantum diagonal computation method. This architecture scheme provides certain circuit-complexity advantages over the amplitude encoding scheme like fewer interacting qubits and low classical overhead for designing the quantum circuit. Hence, GQHT can be used in place of QHT even if we are working with $L2$ normalized input space like for a Quantum centroid-based classifier for exploiting the efficiency of GQHT circuital implementation. Indeed, a quantum algorithm that exploits fewer resources is beneficial. We perform numerical simulations of quantum circuits that establish the workability of the classical information mapping scheme and our generalized algorithm. We use PennyLane, a Python-based quantum machine learning library for numerical simulations \cite{bergholm2018pennylane}.

We organize this paper as follows: In  Section (\ref{sec2}), we discuss the GQHT algorithm and also compare it with the traditional QHT. This discussion will be carried out on the quantum circuital model for implementation of GQHT, highlighting the advantages that arise from its design. In the next Section (\ref{sec3}), we discuss the training and testing of two binary classifiers: the Logistic regression binary classifier and the Centroid-based binary classifier. These classifiers use our GQHT as an important subordinate to get quantum advantages like less space complexity and faster computation. We also evaluate their performance on both public-benchmark datasets: the Iris dataset and the Seeds dataset, and artificial datasets: the Blobs dataset and the Two Half Moons datasets. Finally, we present our concluding remarks in the last section (\ref{section4}).
\section{Generalize Quantum Hadamard Test}
\label{sec2}
\begin{figure}
    \centering
    \includegraphics[ width=10 cm]{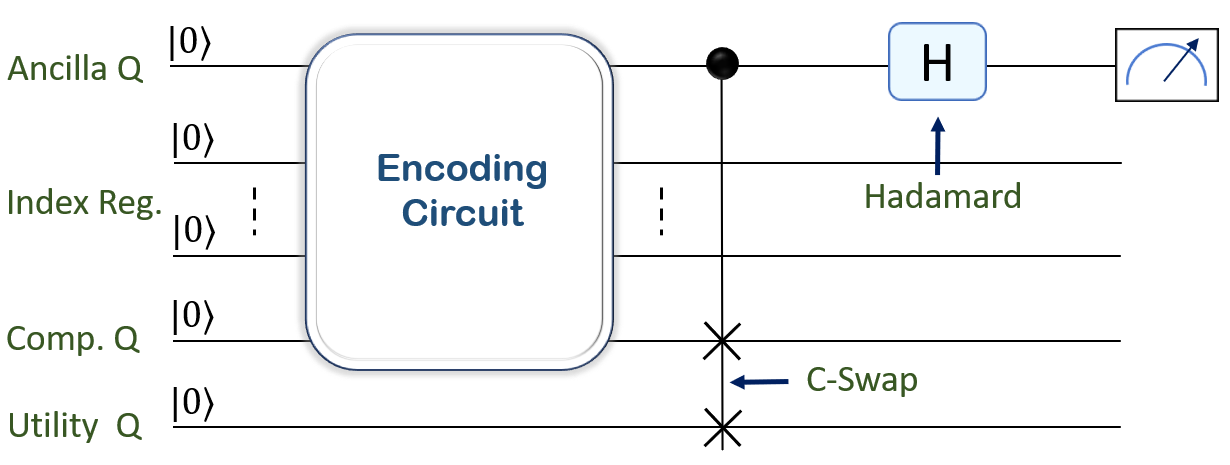}
    \caption{Schematic for the quantum circuit for GQHT algorithm. The abbreviation is used here: Ancilla qubit as Ancilla Q, Index Register as Index Reg., Component qubit as Comp. Q and Utility qubit as Utility Q.  }
    \label{fig1}
\end{figure}
We begin by computing the inner product between two \textit{bound} vectors, $\mathbf{x_p}, \mathbf{x_q} $, which belong to a set of $N$-dimensional real space, $X\subset\mathbb{R}^N$. 
We need to use a quantum circuit-based quantum feature mapping scheme $U_{X}$, that maps this input real space to quantum composite Hilbert space, $\mathcal{H};\text{ } U_{X}:X \rightarrow \mathcal{H}$, where all the input data points are represented as uniform superposition. This composite Hilbert space consists of two subspaces, one subspace corresponds to a single  ancilla qubit, while the other  corresponds to  a quantum register that stores the individual quantum states  as
\begin{equation}
    \label{eq2.1}
    \ket{\mathbf{x}}=\frac{1}{\sqrt{2^n}}\sum_{j=0}^{2^n-1}\ket{j}\left(x_j\ket{0}-\sqrt{1-x_j^2}\ket{1}\right),
\end{equation}
where, $\ket{j}\in\{0,1\}^n$ are the computational basis of a quantum Hilbert space of dimension $2^n$, which consists of $n=\lceil log_2N\rceil $ single qubits, which is used to enumerate the components of the given vector, and the value of their corresponding components, $x_j$ are stored in another single qubit, referred to as \textit{component} qubit. Here, we need to pad the input vectors with zeros when the number of components of the real input space is less than the dimension of quantum Hilbert space, $N<2^n$. The quantum space encoding unitary mapping circuit maps the reference state, $\ket{0}^{\otimes(n+2)}$ to 
\begin{align}
    \ket{\Psi_0}&=U_{X}\ket{0}^{\otimes(n+2)}\notag\\
    &=\frac{1}{\sqrt{2}}\left[\ket{0}\ket{\mathbf{x}_p}+\ket{1}\ket{\mathbf{x}_q}\right]\label{eq2.2}
\end{align}
Once we encode the input space, then we now prepare for the next stage where we transform the quantum state Eq. (\ref{eq2.2}) as
\begin{align}
    \ket{\psi_1}&=H\text{ }C-swap\ket{\Psi_0}\ket{0}, \label{eq2.3}\\
    &=\frac{1}{2}\left[\ket{0}(\ket{\Tilde{\mathbf{x}}_p}+\ket{\Tilde{\mathbf{x}}_q})+\ket{1}(\ket{\Tilde{\mathbf{x}}_p}-\ket{\Tilde{\mathbf{x}}_q})\right].\label{eq2.4}
\end{align}
Here, $H$ and $C-swap$ are Hadamard gate and controlled-swap gate, respectively. We introduce another qubit that we call \emph{utility} qubit in  Eq. (\ref{eq2.3}). We apply a three-qubit  $C-swap$ gate, where ancilla is a control qubit and its target qubits are component qubit and utility qubit, and a single $H$ gate on ancilla qubit in sequence to get the quantum state in Eq. (\ref{eq2.4}). In Eq. (\ref{eq2.4}), two states are
\begin{align}
    \ket{\Tilde{\mathbf{x}}_p}&=\frac{1}{\sqrt{2^n}}\sum_{j=0}^{2^n-1}\ket{j}\left(x_{pj}\ket{0}-\sqrt{1-x_{pj}^2}\ket{1}\right)\ket{0}, \text{ and}\notag\\
    \ket{\Tilde{\mathbf{x}}_q}&=\frac{1}{\sqrt{2^n}}\sum_{j=0}^{2^n-1}\ket{j}\ket{0}\left(x_{qj}\ket{0}-\sqrt{1-x_{qj}^2}\ket{1}\right).\notag
\end{align}
Next, we calculate the expectation value of the Pauli-Z operator acting over the ancilla qubit. The spectral decomposition Pauli-Z operator is a linear composition of computational basis projection operators, $\hat{Z}=\ket{0}\bra{0}-\ket{1}\bra{1}$. Hence, the expectation value of the Pauli-Z operator on the ancilla qubit is
\begin{align}
    \langle\hat{Z}\rangle&= Tr(\ket{\Psi_1}\bra{\Psi_1}\hat{Z})\equiv \langle\Psi_1|\hat{Z}|\Psi_1\rangle\notag\\
    &=|\langle0|\Psi_1\rangle|^2-|\langle1|\Psi_1\rangle|^2\label{eq2.5} \\
    &=\frac{1}{2^n}\langle\mathbf{x}_p,\mathbf{x}_q\rangle, \label{eq2.6} 
\end{align}
where $Tr(.)$ is an trace operation. From Eq. (\ref{eq2.6}), we get the inner product of two vectors, $\langle\mathbf{x}_p,\mathbf{x}_q\rangle$, form our Generalized quantum interference test. Since from Eq. (\ref{eq2.5}), we find that the information of this inner product is concealed in the probability amplitudes of the computational basis of ancilla qubit, therefore we need to run this algorithm from a large but fixed number of times and use statistical inference to get the estimated value of probability amplitude. Fig. (\ref{fig1}) depicts the GQHT algorithm.

\textbf{Comparison with QHT:} We now compare our protocol with the well-known and established quantum similarity-measurement protocol, the quantum Hadamard test. In this protocol, we need to compute the $L2$-norm of each input,  before mapping it to the quantum state; and so, the classical information-embedded quantum state that represents the $L2$-normalized input is,
\begin{equation}\label{eq2.7}
    \ket{\mathbf{x}'}=\sum_{j=0}^{2^n-1}x'_j\ket{j},
\end{equation}
where $(.)'$ denotes that this vector is $L2$-normalized. This quantum feature mapping refers to amplitude encoding, where the components of the normalized vector are stored in the probability amplitude of the basis of $n$-qubit register, therefore it does not require an extra qubit, say component qubit as it is required in our protocol to store the information. It is important to note that amplitude encoding is a linear mapping of input space to quantum Hilbert space, while our quantum feature mapping is nonlinear, Eq. (\ref{eq2.1}). GQHT does not use this extra component $\sqrt{1-x^2_j}$, however, this nonlinear quantum feature mapping allows it to compute inner products beyond $L2$ normalized inputs. Let's start the steps to get the similarity between two vectors, $\ket{\mathbf{x}'}_p$ and $\ket{\mathbf{x}'}_q$, using the quantum Hadamard test. Here, the first step is identical to our first step, as shown in Eq. (\ref{eq2.2}). For the second step,  since we do not require utility qubit and hence there is no need for a controlled-swap gate. Therefore, we only apply a Hadamard gate on the ancilla qubit and then in the final step, measure the expectation value of the Pauli-Z operation on the ancilla qubit, which returns
\begin{equation}
    \label{eq2.8}
    \langle\hat{Z}\rangle=\langle\mathbf{x}'_p,\mathbf{x}'_q\rangle.
\end{equation}
On comparing the basis of qubit requirement, the quantum Hadamard test requires two fewer qubits as well as one less transformation. These benefits also limit its capacity to work with a more generalized representation of inputs as we cover with our protocol. In the realm of quantum machine learning, our protocol can replace the quantum Hadamard test with an additional need of scaling factor $\frac{1}{2^n}$. In a nutshell, we may find that where the quantum Hadamard test is used for quantum machine learning tasks, there our protocol may take its place.  In the following section where we discuss the quantum circuit architecture of GQHT.
\subsection{Quantum Circuit Design for GQHT}
\label{subsec2.1}
\begin{figure}
    \centering
    \includegraphics[ width=13 cm]{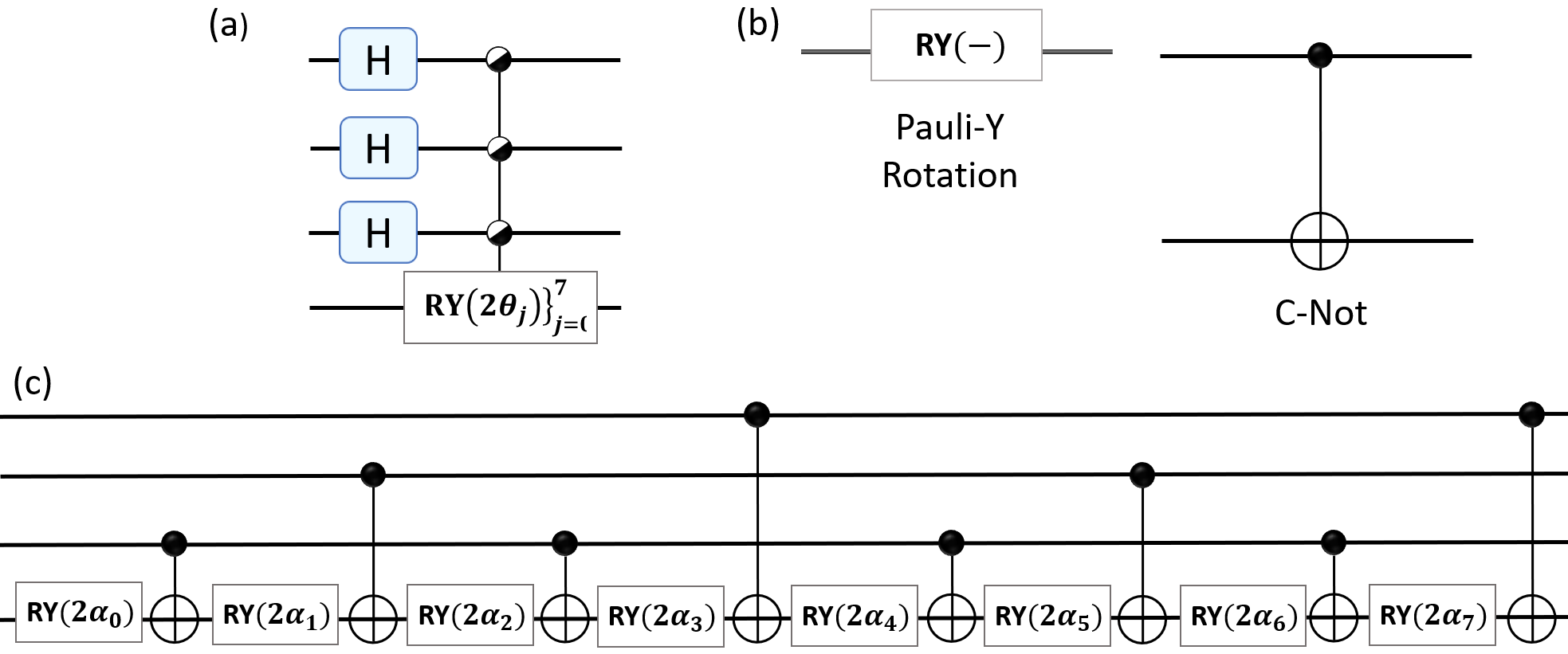}
        \caption{The compact form of the encoding map circuit is shown in figure \textbf{(a)}. In figure \textbf{(b)}, two gates are shown that are required for efficient decomposition. The quantum circuit architecture in figure (\textbf{c}), for efficient decomposition of the 3-qubits uniformly-controlled Pauli-Y rotation gates, is shown in figure (\textbf{a}).  }
    \label{fig2}
\end{figure}
In this subsection, we discuss the design of a quantum circuit for implementing our protocol. It is crucial to construct an efficient circuit using the universal set of gates, since a non-efficient implementation may forbid getting fruitful quantum advantage. The efficient circuit design should minimize the requirement of two-qubit gates ( two-qubit gates since it requires quantum correlation) while keeping the depth of the circuit low. The consequence of efficient circuit design is to save from the risk of quantum noises along with keeping the cost of valuable quantum resources in check.

In our protocol, only the input space encoding unitary map, $U_X$ requires special attention for quantum circuit designing. The quantum state representation of the input space, as shown in Eq. (\ref{eq2.2}), provides the evidence of use of diagonal unitary evolution of the reference state. Hence, the encoding map, $U_X$, requires a diagonal computation encoding circuit scheme. To illustrate this scheme, we design a prototypical quantum circuit. Fig. (\ref{fig2}, a) shows the compressed form of the unitary circuit, $U_X$. From Fig. (\ref{fig2}, a), we see that it consists of two parts: in the first part, it requires the application of Hadamard gates on subsystems, and then needs to apply 3-qubits uniformly-controlled parameterized Pauli-Y rotations in sequence. Here, we need to get the efficient decomposition of this sequence of uniformly controlled gates in terms of single parameterized Pauli-Y rotation gates and two-qubit C-not gates, as shown in Fig (\ref{fig2}, b). 
M\"{o}tt\"{o}nen, \emph{et. al.} \cite{mottonen2004quantum} proposed an efficient protocol to get such decomposition. Such a decomposition for 3-qubits uniformly-controlled rotation gates is shown in Fig. (\ref{fig2}, c). It is noted that we need to apply the angle vector, $\boldsymbol{\theta}$, such that its entries are $\theta_j=arccos(x_j)$, as shown in (\ref{fig2}, a) to represent the quantum state in Eq. (\ref{eq2.2}), but we see from Fig.  (\ref{fig2}, c), we are applying angle vector, $\boldsymbol{\alpha}$. This is because entries of $\alpha_j,j=0,\dots,7$, is a linear combination of components in angle vector $\boldsymbol{\theta}$. The relationship between three vectors is given by the linear transformation,
\begin{equation}
\label{eq2.1.1}
    M\boldsymbol{\alpha}=\boldsymbol{\theta},
\end{equation}
\begin{figure}
\centering
\includegraphics[ width=15 cm]{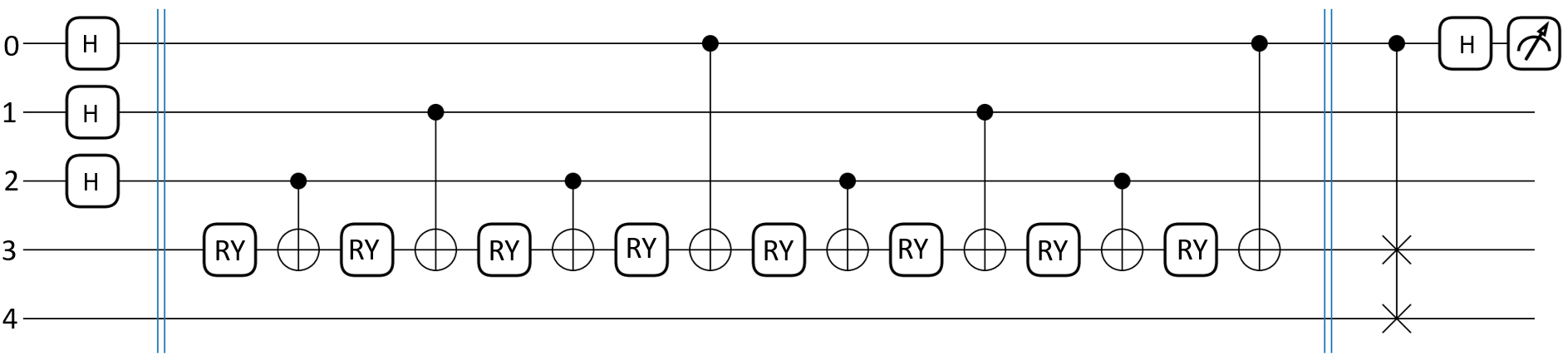}
\caption{This is a quantum circuit to implement the GQHT algorithm for getting the inner product between two vectors having four components. This quantum circuit develops directly in PennyLane during the numerical simulation. Here, the indices of qubits 0, (1,2), 3, and 4 correspond to ancilla qubit, index register, component qubit, and utility qubit, respectively. }
\label{fig3}
\end{figure}
where $M$ is $8\times 8$ orthogonal and unitary matrix. Elements of this matrix, as well as the design of our quantum circuit, are driven from the representation of the computation basis of 3-qubit controlled quantum state systems in terms of Gray code, as shown in Table (\ref{table1}).  Corresponding to the flip of a single bit in successive Gray code order ($g_j$) determines the control-qubit place in the quantum circuit architecture, as illustrated in Table (\ref{table1}) and Fig. (\ref{fig2}, c). Following we get the components of matrix M, as
\begin{equation}
\label{eq2.1.2}
    M_{sq}=\frac{\left(-1\right)^{b_s.g_q}}{2^3},\quad s,q=0, \dots, 7,
\end{equation}
\begin{table}[h]
    \begin{minipage}{0.5\textwidth}
        \centering
        \begin{tabular}{|c|c|c|c|c|}
            \hline
            MSB & MB & LSB & Gray code  & Binary  code \\ \hline
            0 & 0 & 0 & $g_0$ & $b_0$ \\ \hline
            0 & 0 & 1 & $g_1$ & $b_1$ \\ \hline
            0 & 1 & 1 & $g_2$ & $b_3$ \\ \hline
            0 & 1 & 0 & $g_3$ & $b_2$ \\ \hline
            1 & 1 & 0 & $g_4$ & $b_6$ \\ \hline
            1 & 1 & 1 & $g_5$ & $b_7$ \\ \hline
            1 & 0 & 1 & $g_6$ & $b_5$ \\ \hline
            1 & 0 & 0 & $g_7$ & $b_4$ \\ \hline
        \end{tabular}
        \caption{Gray code representation for three-bit strings. Abbreviations used are most significant bit  as MSB, middle bit as MB, and least significant bit as LSB. The first three columns present gray code representation, while the last two columns show gray code order correspondence with binary code.   }
        \label{table1}
    \end{minipage}%
    \hspace{0.01\textwidth}
    \begin{minipage}{0.5\textwidth}
        \centering
        \begin{tabular}{|c|c|c|c|c|c|c|c|c|}
            \hline
            & $g_0\equiv b_0$ &$g_1\equiv b_1$ & $g_2\equiv b_3$ & $g_3\equiv b_2$ & $g_4\equiv b_6$ & $g_5\equiv b_7$ & $g_6\equiv b_5$ & $g_7\equiv b_4$  \\
             \hline
           $b_0$& 1 & 1 & 1 & 1 & 1 & 1 & 1 & 1   \\ \hline
           $b_1$& 1 & -1 & -1 & 1 & 1 & -1 & -1 & 1   \\ \hline
           $b_2$& 1 & 1 & -1 & -1 & -1 & -1 & 1 & 1   \\ \hline
           $b_3$& 1 & -1 & 1 & -1 & -1 & 1 & -1 & 1   \\ \hline
           $b_4$& 1 & 1 & 1 & 1 & -1 & -1 & -1 & -1   \\ \hline
           $b_5$& 1 & -1 & -1 & 1 & -1 & 1 & 1 & 1   \\ \hline
           $b_6$& 1 & 1 & -1 & -1 & 1 & 1 & -1 & -1   \\ \hline
           $b_7$& 1 & -1 & 1 & -1 & 1 & -1 & 1 & -1   \\ \hline
        \end{tabular}
        \caption{ Orthogonal matrix gives the linear relation between $\boldsymbol{\theta}$ and $\boldsymbol{\alpha}$. $\frac{1}{2^3}$ should be multiplied with this matrix. This table is prepared in the following two steps; in step 1, we take $8\times 8$ Hadamard transformation matrix, and in the final step, we permute the columns according to the gray code order correspondence with binary code as shown in the last two columns in Table (\ref{table1}). This correspondence relation between the two codes is also shown in the column header.}
        \label{table2}
    \end{minipage}
\end{table}
where $s \text{ and } q$ are arbitrary indices, and $b_s.g_q$ is element-wise binary dot product. Matrix M is shown in Table (\ref{table2}) is the required linear transform as shown in Eq. (\ref{eq2.1.1}). It is a real matrix that belongs to an orthogonal class, and hence its inverse $M^{-1}$ is equal to its transpose, $M^T$. Fianlly, we compute the $\boldsymbol{\alpha}$ in linear combination of $\boldsymbol{\theta}$ using, $\boldsymbol{\alpha}=M^T \boldsymbol{\theta}$. In addition, we need not compute each element explicitly using Eq. (\ref{eq2.1.2}). However, we use a trick that cancels the requirement to explicitly compute each component of this matrix. The trick is that we permute the columns of the corresponding dimension Hadamard gate according to the order correspondence between gray and binary codes. This order-correspondence can be seen in columns $IV \text{ and } V$ of Table (\ref{table1}) and correspondingly columns-wise permuted matrix is seen in Table (\ref{table2}). Both mathematical tricks, regarding the matrix inverse and preparation of the matrix, minimize the classical computing cost of getting $\boldsymbol{\alpha}$ in terms of $\boldsymbol{\theta}$. Additionally, to further reduce computation, we avoid repeatedly computing the matrix product  $M^T \boldsymbol{\theta}$. Instead, we prepare a function that performs the matrix product for a given matrix $M^T$  and variable vector $\boldsymbol{\theta}$. This function takes $\boldsymbol{\theta}$ as an argument and returns the vector $\boldsymbol{\alpha}$.

Now, we discuss the advantages of our quantum circuit scheme concerning the amplitude-encode quantum states required for implementing the QHT algorithm \cite{araujo2021divide}. These advantages are as follows:
\begin{enumerate}
    \item It requires a less-connected quantum register. From Fig. (\ref{fig2}), we conclude that a large number of qubits need to be connected with a single qubit, rather than to each other.
    \item The depth of our quantum circuit architecture is low and consequently low risk of quantum decoherence. Our quantum encoding circuit's depth is $\mathcal{O}(N)$ to map any $N$-dimensional input space. The depth of our quantum circuit architecture is identical to the quantum circuit for amplitude-encode state preparation. However, it requires more classical overhead since we first need to compute required angles, as discussed in algorithm 1 of this paper \cite{araujo2021divide} and then we need to use  M\"{o}tt\"{o}nen, \emph{et. al.} \cite{mottonen2004quantum} scheme to deploy uniformly controlled rotation gates.
   \item Whenever required, we may individually encode the information of vectors. See appendix (\ref{sec5.1}), where we encode the training dataset and test point individually. This approach minimizes the requirement of classical overhead as well as quantum resources. However, we are not able to use this method to prepare amplitude-encoded quantum states. 
\end{enumerate}
It is worth mentioning that GQHT has a different quantum feature mapping protocol for mapping the input space. This quantum feature mapping scheme can be used as quantum feature mapping for variational classifiers \cite{schuld2020circuit}. 

Finally, we perform a numerical simulation of our circuit using \emph{PennyLane}, a Python-based library for the numerical simulation of quantum circuits. Here, we compute scaled-inner product two vectors, $(0.1,0.25,-1,0.9) \text{ and } (-1,0.75,0.65,0.89)$ using our protocol. The required quantum circuit is shown in Fig. (\ref{fig3}). The result that we get after implementing this circuit is the required scaled-inner product between the given two vectors, $0.06$. In the appendix (\ref{sec5.1}), we briefly discuss the simultaneous computation of the inner product between the number of pairs of vectors. Next, when our vectors are binary-valued, they can be used directly to implement a quantum circuit model of an artificial neuron that is discussed by Tacchino \emph{et. al.} \cite{tacchino2019artificial}.

\section{Machine Learning Classifiers}
\label{sec3}
\begin{figure}
    \centering
    \includegraphics[ width=10 cm]{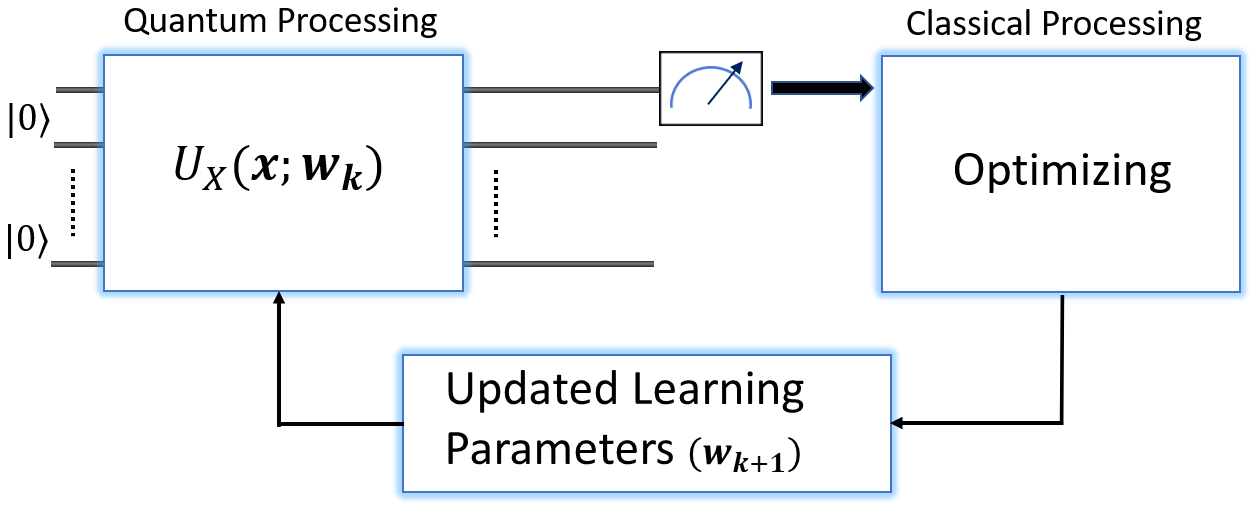}
    \caption{It is a schematic of training of Logistic regression binary classifier. We use the quantum processing unit to run the GQHT algorithm that returns the inner product between inputs and learning parameters, then its result feeds into the classical processing unit for learning parameters optimization. It is an iterative process, therefore after each iteration, we get the updated learning parameters that again feed into the quantum processing unit.   }
    \label{fig4}
\end{figure}
It is impressive to show the working of our proposed Generalized test. It is used as the important subordinate of two classical machine learning classifiers: Logistic Regression binary classifier and Centroid-based binary classifier. Noted that integration of our quantum algorithm-based subordinate with required classical subordinates for the given classifiers put these on the same breed of machine learning models that follow the hybrid quantum-classical approach. The interesting aspect of our hybrid classifiers is that they have inherent basic characteristics like trainability and generalization from their corresponding classical counterparts. Our approach is underlying quantum-enhanced machine learning approaches where we are provided with quantum subordinates in view of getting computational speedups for established machine learning algorithms \cite{dunjko2016quantum, rebentrost2014quantum, lloyd2014quantum}. 
\begin{figure}
    \centering
    \includegraphics[ width=12.5 cm]{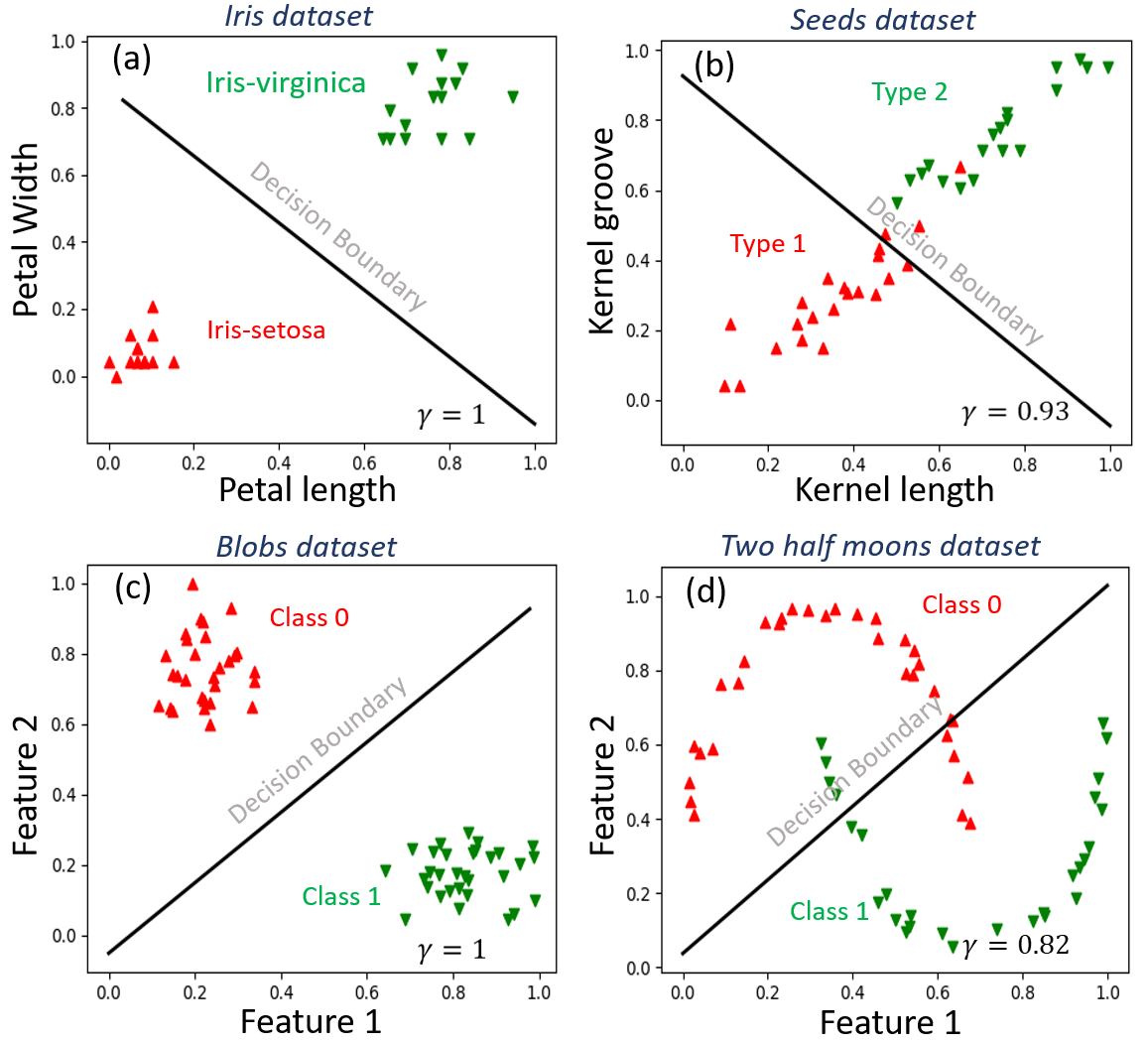}
    \caption{Decision boundary plots over four datasets are shown.  Names of datasets and their generalization accuracies are mentioned above over subfigures and the lower-right side position of plots, respectively.  In all plots, only test inputs are shown.  }
    \label{fig5}
\end{figure}

We now discuss the binary classification problem setup. Consider we have labelled a data set $\mathcal{D}=\{(\mathbf{x}^m, y^m)\}_{m=0}^{M-1}\in (X,\{0,1\}),$ consists of $M$ inputs that live in $N$-dimensional real  inner product space, $X\subset\mathbb{R}^N$. This labeled data set is split into two sets, as the training set ($\mathcal{T})$ and test set ($\mathcal{S}$). One part will be used for training during the training stage and the other part will be used to evaluate the performance of the trained model during the testing stage. A learning model is preferable when it performs well at the testing stage, i.e. on generalization. 

\subsection{Logistic Regression Binary Classifier}
\label{subsec3.1}
The Logistic regression binary classifier belongs to the class of parameterized machine learning algorithms that follow the approach of machine learning design as an optimization problem. We now describe its main components: hypothesis space, objective function, decision function, and performance analysis. Let's describe its first component, hypothesis space:
\begin{equation}\label{eq3.1.1}
    h(\mathbf{x}^m; \mathbf{w}, w_0)=\sigma(\langle\mathbf{w},\mathbf{x}^m\rangle+w_0),
\end{equation}
where $\mathbf{x}^m\in X$, $(\mathbf{w}, w_0)$ constitutes the learning parameters and bias, respectively that can be exploited to search the hypothesis space during optimization. The sigmoidal function, $\sigma$ is defined as: 
\begin{equation*}
    \sigma(z)=\frac{1}{1+e^{-z}},
\end{equation*}
where $z$ is an arbitrary variable. For our work, we add one more dimension in the input space by providing an additional component $x_{m0}=1$ to each input, such that the equation in parenthesis of Eq. (\ref{eq3.1.1}) is rewritten as $\langle\mathbf{w},\mathbf{x}^i\rangle+w_0x_{m0}$. This inner product can be efficiently computed from our Generalized quantum interference test, while other calculations are carried out by classical processors. Once we come up with the hypothesis space, then we need to define the objective function,
\begin{equation}
    \label{eq3.1.2}
    J(\mathbf{w},w_0,\mathcal{T})=\frac{1}{B}\sum_{b=0}^{B-1}\mathcal{L}( h(\mathbf{x}^b; \mathbf{w}, w_0), y^b),
\end{equation}
where $B$ is a randomly selected batch size of training inputs. We compute total loss over this batch inputs via the loss function, $\mathcal{L}()$. Here, a binary cross-entropy loss function is selected. Next, we select the optimizer which iteratively minimizes the objective function for the given number of times. At the end of this step, we come up with the optimized parameter value $\mathbf{w}^*,w_0^*$. We choose the Stochastic batch gradient descent as our optimizer. Finally, we need to design the decision function as 
\begin{equation}
    y'^{m}=\begin{cases}
        1,\qquad  \sigma(h(\mathbf{x}^m; \mathbf{w}^*, w_0^*))>0.5 \\
        0, \qquad \text{otherwise}.
    \end{cases}
\end{equation}
To check the performance of our learning model on generalization, we conduct performance analysis over the test set, $\mathcal{S}$. We run this algorithm in four datasets: the first two belong to public-benchmark datasets such as the Iris and Seeds datasets that are shown in the first row of Fig.(\ref{fig5}), while the last two datasets are artificial datasets like Blobs and two half moons dataset, shown in the second row of Fig. (\ref{fig5}). Here, we use basic Python code to run this classifier. We use $70\%$ of data for training and the remaining instances for testing. 
\begin{figure}
    \centering
    \includegraphics[width=9 cm]{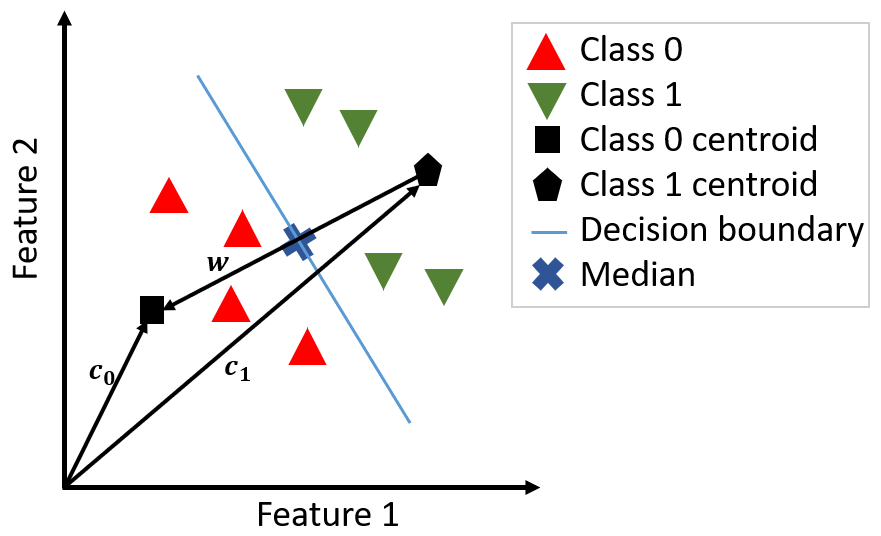}
    \caption{Schematic of Centroid-based binary classifier. We can see how two terms, median value $\mathbf{c}$ and  $\mathbf{w}$ that is the difference between centroid-vectors $\mathbf{c_0}\text{ and }\mathbf{c_1}$ that decides the orientation of the decision boundary. }
    \label{fig6}
\end{figure}

It is important to highlight certain advantages of using GQHT as a subordinate in this classifier. Firstly, in comparison to classical machines, we minimize the space complexity exponentially, i.e., we need only $\mathcal{O}(log_2 N)$, the number of qubits for representing any arbitrary $N$-dimensional vector; whereas in the classical machine, we need given number $2^N$ in numbers for storing the indices of the vector plus additional binary strings to store their corresponding components with a given precision. Secondly, we discuss regarding the statistical learning perspective of our GQHT scheme. From Fig. (\ref{fig5}), we find that we get the decision function in the Min-max pre-processed input space. There is no change in the distribution of input space due to Min-max normalization. Our Logistic regression binary classifier is inspired by the variational class of quantum classifiers, where the learning parameters of parameterized quantum circuits are iteratively optimized by classical processors. Implicitly, we can say that we propose a variational classifier that works on other than  $L2$ normalized input space. On the other hand, we also find that the working of this classifier is identical to an artificial neuron, and hence we can exploit it to formalize a hybrid quantum-classical quantum neural network, as discussed by Tacchino \emph{et. al.} \cite{tacchino2020quantum}.
\subsection{Cetroid-based Binary Classifier}
We adopt an interesting approach to designing a centroid-based binary classifier model that is different from the conventional centroid-based binary classifier model. This approach is driven from \emph{Learning with Kernels} book \cite{scholkopf2018learning}. The decision equation for assigning the label $y$ of test data point  $\mathbf{x}$ by this model is,
\begin{equation}
    \label{eq3.2.1}
    y=sgn\langle(\mathbf{x}-\mathbf{c},\mathbf{w}\rangle,
\end{equation}
where two terms are elaborated as: $\mathbf{c}:=(\mathbf{c}_0+\mathbf{c}_1)/2$ is the mean of the centroids and $\mathbf{w}:=(\mathbf{c}_0-\mathbf{c}_1)$ is difference between centroids ( mean is synonyms of centroid). Here, $sgn(.)$ is a step function that returns the class label according to the sign of \emph{arc cosine} of value under parenthesis of Eq. (\ref{eq3.2.1}). Centroids for class 0 and class 1 are given, respectively, as 
\begin{align}
    \mathbf{c}_0&=\frac{1}{M}\sum_{\{m|y^m=0\}}\mathbf{x}^m \label{eq3.2.2} \\
    \mathbf{c}_1&=\frac{1}{M}\sum_{\{m|y^m=1\}}\mathbf{x}^m \label{eq3.2.3}
\end{align}
where $M$ is the number of samples in class $0$ (class $1$), i.e., our training data points are balanced. When we put the Eqs. (\ref{eq3.2.2}) and (\ref{eq3.2.3}) in Eq. (\ref{eq3.2.1}), we get the equation
\begin{figure}
    \centering
    \includegraphics[height=12.5 cm, width=12.5 cm]{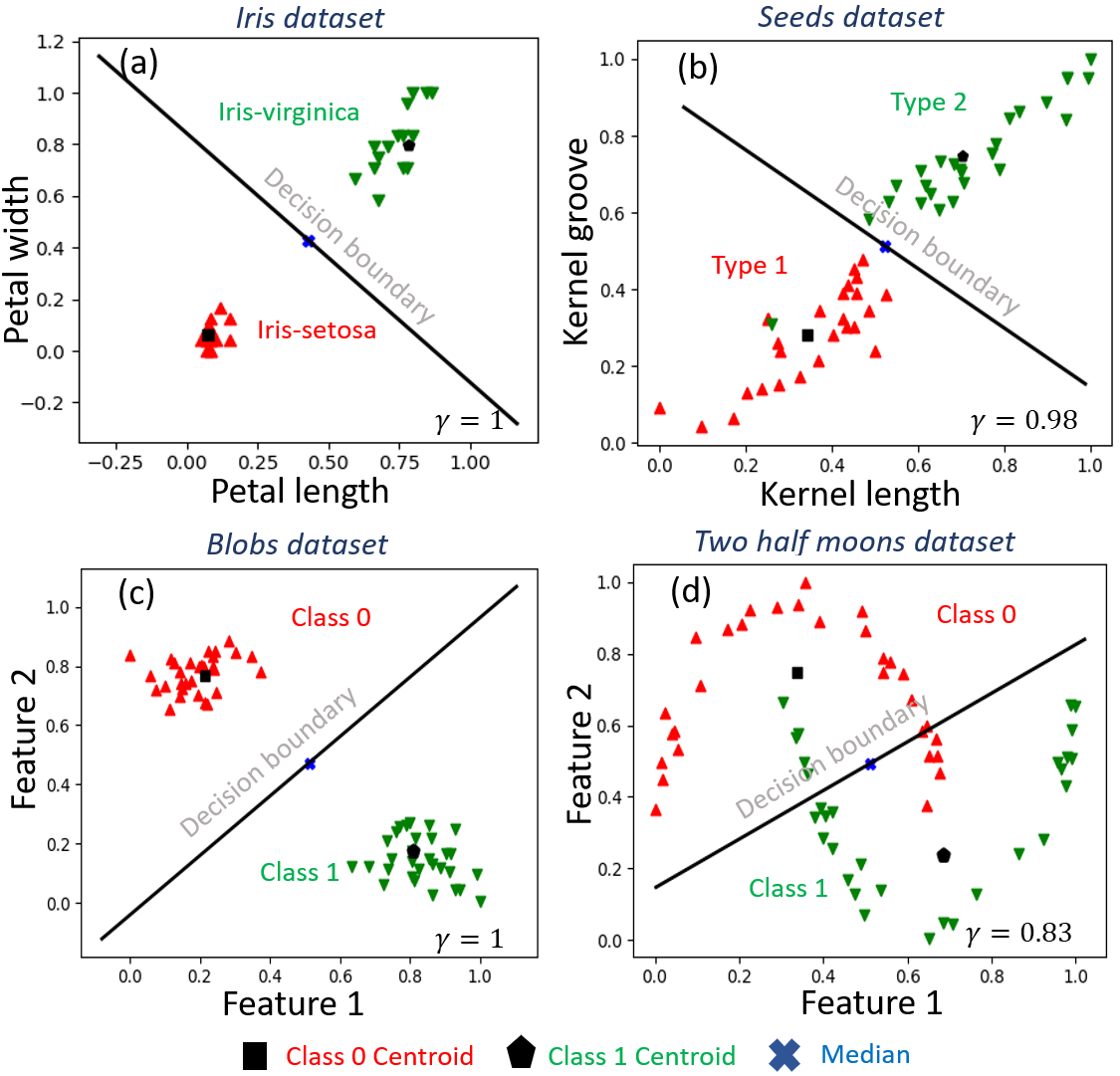}
    \caption{Decision boundary plots over four datasets are shown. The names of the dataset and their corresponding generalization accuracy are shown as titles over subfigures and as legends in the lower-right side of plots, respectively. Other notations like class centroids and medians are presented in the lower bar of the figure.  In all plots, only test inputs are shown.}
    \label{fig7}
\end{figure}
\begin{equation}
\label{eq3.2.4}
    y=sgn\left(\frac{1}{M}\sum_{\{m|y^m=0\}}\langle\mathbf{x}^m,\mathbf{x}\rangle-\frac{1}{M}\sum_{\{m|y^m=1\}}\langle\mathbf{x}^m,\mathbf{x}\rangle+b\right),
\end{equation}
where the offset, $b$ is defined as:
\begin{equation}
    \label{eq3.2.5}
    b=\frac{1}{2}\left(\frac{1}{M^2}\sum_{\{(m,\vec m)|y^m,y^{\vec m}=0\}}\langle\mathbf{x}^m,\mathbf{x}^{ \vec m}\rangle-\frac{1}{M^2}\sum_{\{(m,\vec m)|y^m,y^{\vec m}=1\}}\langle\mathbf{x}^m,\mathbf{x}^{\vec m}\rangle\right).
\end{equation}
From Eqs. (\ref{eq3.2.4}) and (\ref{eq3.2.5}), we find that all the terms are redefined in terms of inner products between test data points to all training data points as in Eq. (\ref{eq3.2.4}) or one training data point to all other training data points that belongs to the same class as in Eq. (\ref{eq3.2.5}). These inner products can be implemented in one go due to quantum parallelism (see appendix (\ref{sec5.1})).   

As the earlier classifier discussed in subsection (\ref{subsec3.1}), we again work with the same datasets, as shown in Fig. (\ref{fig7}). Here, again we perform the Python-based code to do numerical simulation which returns inner products and these are used to get the class of test inputs. For this classifier, a balanced dataset is required. Except for the seeds dataset, other datasets are balanced. To address this, we balance the seeds dataset by selecting an equal number of inputs from both classes during training. 

Finally, we talk about the merit of GQHT concerning this classifier. Firstly, it is important to note that this is not a computationally efficient approach, however, we are taking this in particular to exhibit the strength of our protocol to do a classification task when we work with only bounded input space. Here, we find that the model requirement of this classifier demands the inner product between bounded vectors. This is the merit of our algorithm which may work for this requirement along with other tasks where the QHT algorithm is already used like quantum centroid-based classifiers. Another important point is that when we integrate our GQHT algorithm for the quantum centroid-based quantum classifier, we can minimize the quantum resources and classical calculation overhead during quantum circuit implementation.
\section{Conclusions}
\label{section4}
We design the Generalized version of the amplitude-mapped quantum Hadamard test. In addition, we discuss the efficient quantum circuit scheme for implementing this algorithm. We also validate the design of this scheme by numerical simulation of a prototype quantum circuit by using PennyLane, a Python-based quantum computation library. We further demonstrated the application of this scheme by incorporating it into two binary classifiers: Logistic regression binary classifier and Centroid-based binary classifier; and solving classification learning problems over four datasets. 
In addition, our GQHT may be applied in the field of quantum neural networks. Tacchino \emph{et. al.} \cite{tacchino2019artificial} presented the quantum circuital implementation of a binary-valued artificial neuron. They present the hypergraph-based implementation of this neuron which requires a lengthy classical backend. We can directly replace their quantum computational model with our GQHT. There is another artificial neuron quantum computational implementation method that is discussed by  Yan \emph{et. al.} \cite{yan2020nonlinear}, and our GQHT can replace the part of the algorithm where the inner product is computed. In summary, our GQHT algorithm shows great promise for a broad range of machine learning applications and its full potential can be further explored in future works.

\section{Appendix}
\subsection{GQHT getting Inner Product between the Training set and a Test point}
\label{sec5.1}
\begin{figure}
    \centering
    \includegraphics[ width=15cm]{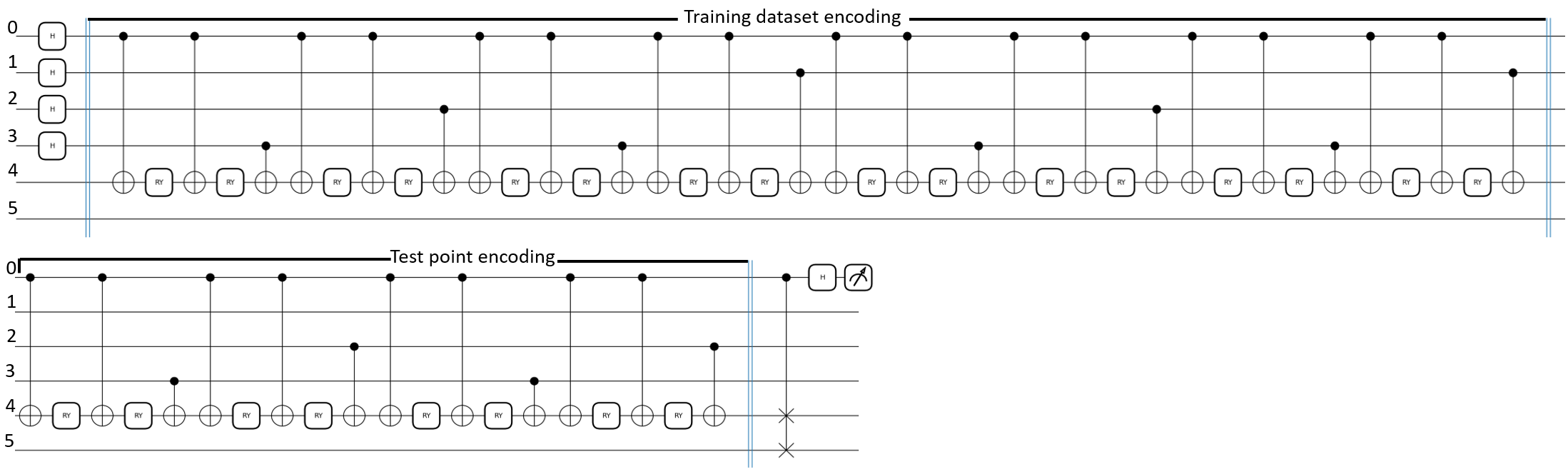}
    \caption{Quantum circuit schematic for getting the inner product between the training dataset and the test point. It is generated from the PennyLane.  Index 0, 1, (2, 3), 4, and 5 correspond to ancilla qubit, sample index qubit, component index register, component qubit, and utility qubit, respectively. In the training dataset encoding section, we need to apply a 0 basis controlled Pauli-Y rotation gate, so it can be implemented by applying this sequence of gates $C-NotR_y(\theta)C-NotR_y(\theta)$, where $\theta$ is a parameter. While in the test point encoding, we need to apply 1 basis controlled Pauli-Y rotation gate, and it can be implemented with the same sequence of the gates as discussed earlier with a single change is that the first single qubit Pauli-Y rotation is replaced by its Hermitian conjugate version.   }
    \label{fig8}
\end{figure}
We compute the inner product of a training set, say $\mathcal{D}=\{\mathbf{x}_m\}$ consists of $M$ samples and a test point $\Tilde{\mathbf{x}}$ simultaneously, using our GQHT algorithm. First, we need to represent the training dataset in a quantum state as
\begin{equation}
    \ket{\mathcal{D}}=\frac{1}{\sqrt{2^{(p+n)}}}\sum_{m=0}^{2^p-1}\ket{m}\sum_{j=0}^{2^n-1}\ket{j}\left(x_{mj}\ket{0}-\sqrt{1-x_{mj}^2}\ket{1}\right),
\end{equation}
where the index of samples is stored in the first quantum register consisting of  $p=\lceil log_2M\rceil$ qubits, and the test point is stored as in Eq. (\ref{eq2.1}). Then, we need to apply the GQHT algorithm as discussed in section (\ref{sec2}). When we do get the Pauli-Z operator expectation value over the ancilla qubit, we get
\begin{equation}
    \langle\hat{Z}\rangle=\frac{1}{2^{(p+m)}}\sum_{m=0}^{2^p-1}\langle\mathbf{x}_m,\Tilde{\mathbf{x}}\rangle.
\end{equation}
In Fig. (\ref{fig8}), we illustrate a prototype where the training dataset consists of two vectors, vector 1 $(1.0, 0.25, -0.36, -0.98)$ and vector 2 $(-0.1, 0.37, 0.65, 0.45)$ and the test point is $(0.75, 0.1, 0.25, 0.25)$. The final result comes after the implementation of the quantum circuit is $0.069$. What is important with quantum circuit implementation for this task is that we can encode the training dataset and test dataset \emph{separately}. It is interesting since it can not be possible when we try to encode information in the probability amplitude of the quantum system, which is required for the QHT. It supports us to minimize the load of classical pre-processing.


\begin{thebibliography}{32}
\ifx \bisbn   \undefined \def \bisbn  #1{ISBN #1}\fi
\ifx \binits  \undefined \def \binits#1{#1}\fi
\ifx \bauthor  \undefined \def \bauthor#1{#1}\fi
\ifx \batitle  \undefined \def \batitle#1{#1}\fi
\ifx \bjtitle  \undefined \def \bjtitle#1{#1}\fi
\ifx \bvolume  \undefined \def \bvolume#1{\textbf{#1}}\fi
\ifx \byear  \undefined \def \byear#1{#1}\fi
\ifx \bissue  \undefined \def \bissue#1{#1}\fi
\ifx \bfpage  \undefined \def \bfpage#1{#1}\fi
\ifx \blpage  \undefined \def \blpage #1{#1}\fi
\ifx \burl  \undefined \def \burl#1{\textsf{#1}}\fi
\ifx \doiurl  \undefined \def \doiurl#1{\url{https://doi.org/#1}}\fi
\ifx \betal  \undefined \def \betal{\textit{et al.}}\fi
\ifx \binstitute  \undefined \def \binstitute#1{#1}\fi
\ifx \binstitutionaled  \undefined \def \binstitutionaled#1{#1}\fi
\ifx \bctitle  \undefined \def \bctitle#1{#1}\fi
\ifx \beditor  \undefined \def \beditor#1{#1}\fi
\ifx \bpublisher  \undefined \def \bpublisher#1{#1}\fi
\ifx \bbtitle  \undefined \def \bbtitle#1{#1}\fi
\ifx \bedition  \undefined \def \bedition#1{#1}\fi
\ifx \bseriesno  \undefined \def \bseriesno#1{#1}\fi
\ifx \blocation  \undefined \def \blocation#1{#1}\fi
\ifx \bsertitle  \undefined \def \bsertitle#1{#1}\fi
\ifx \bsnm \undefined \def \bsnm#1{#1}\fi
\ifx \bsuffix \undefined \def \bsuffix#1{#1}\fi
\ifx \bparticle \undefined \def \bparticle#1{#1}\fi
\ifx \barticle \undefined \def \barticle#1{#1}\fi
\bibcommenthead
\ifx \bconfdate \undefined \def \bconfdate #1{#1}\fi
\ifx \botherref \undefined \def \botherref #1{#1}\fi
\ifx \url \undefined \def \url#1{\textsf{#1}}\fi
\ifx \bchapter \undefined \def \bchapter#1{#1}\fi
\ifx \bbook \undefined \def \bbook#1{#1}\fi
\ifx \bcomment \undefined \def \bcomment#1{#1}\fi
\ifx \oauthor \undefined \def \oauthor#1{#1}\fi
\ifx \citeauthoryear \undefined \def \citeauthoryear#1{#1}\fi
\ifx \endbibitem  \undefined \def \endbibitem {}\fi
\ifx \bconflocation  \undefined \def \bconflocation#1{#1}\fi
\ifx \arxivurl  \undefined \def \arxivurl#1{\textsf{#1}}\fi
\csname PreBibitemsHook\endcsname

\bibitem[\protect\citeauthoryear{Biamonte et~al.}{2017}]{biamonte2017quantum}
\begin{barticle}
\bauthor{\bsnm{Biamonte}, \binits{J.}},
\bauthor{\bsnm{Wittek}, \binits{P.}},
\bauthor{\bsnm{Pancotti}, \binits{N.}},
\bauthor{\bsnm{Rebentrost}, \binits{P.}},
\bauthor{\bsnm{Wiebe}, \binits{N.}},
\bauthor{\bsnm{Lloyd}, \binits{S.}}:
\batitle{Quantum machine learning}.
\bjtitle{Nature}
\bvolume{549}(\bissue{7671}),
\bfpage{195}--\blpage{202}
(\byear{2017})
\end{barticle}
\endbibitem

\bibitem[\protect\citeauthoryear{Schuld and Petruccione}{2021}]{schuld2021machine}
\begin{botherref}
\oauthor{\bsnm{Schuld}, \binits{M.}},
\oauthor{\bsnm{Petruccione}, \binits{F.}}:
Machine learning with quantum computers
(2021)
\end{botherref}
\endbibitem

\bibitem[\protect\citeauthoryear{Rebentrost et~al.}{2014}]{rebentrost2014quantum}
\begin{barticle}
\bauthor{\bsnm{Rebentrost}, \binits{P.}},
\bauthor{\bsnm{Mohseni}, \binits{M.}},
\bauthor{\bsnm{Lloyd}, \binits{S.}}:
\batitle{Quantum support vector machine for big data classification}.
\bjtitle{Physical Review Letters}
\bvolume{113}(\bissue{13}),
\bfpage{130503}
(\byear{2014})
\end{barticle}
\endbibitem

\bibitem[\protect\citeauthoryear{Lloyd et~al.}{2014}]{lloyd2014quantum}
\begin{barticle}
\bauthor{\bsnm{Lloyd}, \binits{S.}},
\bauthor{\bsnm{Mohseni}, \binits{M.}},
\bauthor{\bsnm{Rebentrost}, \binits{P.}}:
\batitle{Quantum principal component analysis}.
\bjtitle{Nature Physics}
\bvolume{10}(\bissue{9}),
\bfpage{631}--\blpage{633}
(\byear{2014})
\end{barticle}
\endbibitem

\bibitem[\protect\citeauthoryear{Li et~al.}{2015}]{li2015experimental}
\begin{barticle}
\bauthor{\bsnm{Li}, \binits{Z.}},
\bauthor{\bsnm{Liu}, \binits{X.}},
\bauthor{\bsnm{Xu}, \binits{N.}},
\bauthor{\bsnm{Du}, \binits{J.}}:
\batitle{Experimental realization of a quantum support vector machine}.
\bjtitle{Physical Review Letters}
\bvolume{114}(\bissue{14}),
\bfpage{140504}
(\byear{2015})
\end{barticle}
\endbibitem

\bibitem[\protect\citeauthoryear{Schuld et~al.}{2016}]{schuld2016prediction}
\begin{barticle}
\bauthor{\bsnm{Schuld}, \binits{M.}},
\bauthor{\bsnm{Sinayskiy}, \binits{I.}},
\bauthor{\bsnm{Petruccione}, \binits{F.}}:
\batitle{Prediction by linear regression on a quantum computer}.
\bjtitle{Physical Review A}
\bvolume{94}(\bissue{2}),
\bfpage{022342}
(\byear{2016})
\end{barticle}
\endbibitem

\bibitem[\protect\citeauthoryear{Schuld et~al.}{2017}]{schuld2017implementing}
\begin{barticle}
\bauthor{\bsnm{Schuld}, \binits{M.}},
\bauthor{\bsnm{Fingerhuth}, \binits{M.}},
\bauthor{\bsnm{Petruccione}, \binits{F.}}:
\batitle{Implementing a distance-based classifier with a quantum interference circuit}.
\bjtitle{Europhysics Letters}
\bvolume{119}(\bissue{6}),
\bfpage{60002}
(\byear{2017})
\end{barticle}
\endbibitem

\bibitem[\protect\citeauthoryear{Blank et~al.}{2022}]{blank2022compact}
\begin{barticle}
\bauthor{\bsnm{Blank}, \binits{C.}},
\bauthor{\bsnm{Da~Silva}, \binits{A.J.}},
\bauthor{\bsnm{Albuquerque}, \binits{L.P.}},
\bauthor{\bsnm{Petruccione}, \binits{F.}},
\bauthor{\bsnm{Park}, \binits{D.K.}}:
\batitle{Compact quantum kernel-based binary classifier}.
\bjtitle{Quantum Science and Technology}
\bvolume{7}(\bissue{4}),
\bfpage{045007}
(\byear{2022})
\end{barticle}
\endbibitem

\bibitem[\protect\citeauthoryear{Das et~al.}{2023}]{das2023quantum}
\begin{barticle}
\bauthor{\bsnm{Das}, \binits{S.}},
\bauthor{\bsnm{Zhang}, \binits{J.}},
\bauthor{\bsnm{Martina}, \binits{S.}},
\bauthor{\bsnm{Suter}, \binits{D.}},
\bauthor{\bsnm{Caruso}, \binits{F.}}:
\batitle{Quantum pattern recognition on real quantum processing units}.
\bjtitle{Quantum Machine Intelligence}
\bvolume{5}(\bissue{1}),
\bfpage{16}
(\byear{2023})
\end{barticle}
\endbibitem

\bibitem[\protect\citeauthoryear{Schuld et~al.}{2015}]{schuld2015simulating}
\begin{barticle}
\bauthor{\bsnm{Schuld}, \binits{M.}},
\bauthor{\bsnm{Sinayskiy}, \binits{I.}},
\bauthor{\bsnm{Petruccione}, \binits{F.}}:
\batitle{Simulating a perceptron on a quantum computer}.
\bjtitle{Physics Letters A}
\bvolume{379}(\bissue{7}),
\bfpage{660}--\blpage{663}
(\byear{2015})
\end{barticle}
\endbibitem

\bibitem[\protect\citeauthoryear{Mangini et~al.}{2020}]{mangini2020quantum}
\begin{barticle}
\bauthor{\bsnm{Mangini}, \binits{S.}},
\bauthor{\bsnm{Tacchino}, \binits{F.}},
\bauthor{\bsnm{Gerace}, \binits{D.}},
\bauthor{\bsnm{Macchiavello}, \binits{C.}},
\bauthor{\bsnm{Bajoni}, \binits{D.}}:
\batitle{Quantum computing model of an artificial neuron with continuously valued input data}.
\bjtitle{Machine Learning: Science and Technology}
\bvolume{1}(\bissue{4}),
\bfpage{045008}
(\byear{2020})
\end{barticle}
\endbibitem

\bibitem[\protect\citeauthoryear{Rebentrost et~al.}{2018}]{rebentrost2018quantum}
\begin{barticle}
\bauthor{\bsnm{Rebentrost}, \binits{P.}},
\bauthor{\bsnm{Bromley}, \binits{T.R.}},
\bauthor{\bsnm{Weedbrook}, \binits{C.}},
\bauthor{\bsnm{Lloyd}, \binits{S.}}:
\batitle{Quantum hopfield neural network}.
\bjtitle{Physical Review A}
\bvolume{98}(\bissue{4}),
\bfpage{042308}
(\byear{2018})
\end{barticle}
\endbibitem

\bibitem[\protect\citeauthoryear{Zhao et~al.}{2019}]{zhao2019building}
\begin{barticle}
\bauthor{\bsnm{Zhao}, \binits{J.}},
\bauthor{\bsnm{Zhang}, \binits{Y.-H.}},
\bauthor{\bsnm{Shao}, \binits{C.-P.}},
\bauthor{\bsnm{Wu}, \binits{Y.-C.}},
\bauthor{\bsnm{Guo}, \binits{G.-C.}},
\bauthor{\bsnm{Guo}, \binits{G.-P.}}:
\batitle{Building quantum neural networks based on a swap test}.
\bjtitle{Physical Review A}
\bvolume{100}(\bissue{1}),
\bfpage{012334}
(\byear{2019})
\end{barticle}
\endbibitem

\bibitem[\protect\citeauthoryear{Shao}{2020}]{shao2020data}
\begin{barticle}
\bauthor{\bsnm{Shao}, \binits{C.}}:
\batitle{Data classification by quantum radial-basis-function networks}.
\bjtitle{Physical Review A}
\bvolume{102}(\bissue{4}),
\bfpage{042418}
(\byear{2020})
\end{barticle}
\endbibitem

\bibitem[\protect\citeauthoryear{Yan et~al.}{2020}]{yan2020nonlinear}
\begin{barticle}
\bauthor{\bsnm{Yan}, \binits{S.}},
\bauthor{\bsnm{Qi}, \binits{H.}},
\bauthor{\bsnm{Cui}, \binits{W.}}:
\batitle{Nonlinear quantum neuron: A fundamental building block for quantum neural networks}.
\bjtitle{Physical Review A}
\bvolume{102}(\bissue{5}),
\bfpage{052421}
(\byear{2020})
\end{barticle}
\endbibitem

\bibitem[\protect\citeauthoryear{Benedetti et~al.}{2019}]{benedetti2019parameterized}
\begin{barticle}
\bauthor{\bsnm{Benedetti}, \binits{M.}},
\bauthor{\bsnm{Lloyd}, \binits{E.}},
\bauthor{\bsnm{Sack}, \binits{S.}},
\bauthor{\bsnm{Fiorentini}, \binits{M.}}:
\batitle{Parameterized quantum circuits as machine learning models}.
\bjtitle{Quantum Science and Technology}
\bvolume{4}(\bissue{4}),
\bfpage{043001}
(\byear{2019})
\end{barticle}
\endbibitem

\bibitem[\protect\citeauthoryear{Mitarai et~al.}{2018}]{mitarai2018quantum}
\begin{barticle}
\bauthor{\bsnm{Mitarai}, \binits{K.}},
\bauthor{\bsnm{Negoro}, \binits{M.}},
\bauthor{\bsnm{Kitagawa}, \binits{M.}},
\bauthor{\bsnm{Fujii}, \binits{K.}}:
\batitle{Quantum circuit learning}.
\bjtitle{Physical Review A}
\bvolume{98}(\bissue{3}),
\bfpage{032309}
(\byear{2018})
\end{barticle}
\endbibitem

\bibitem[\protect\citeauthoryear{Havl{\'\i}{\v{c}}ek et~al.}{2019}]{havlivcek2019supervised}
\begin{barticle}
\bauthor{\bsnm{Havl{\'\i}{\v{c}}ek}, \binits{V.}},
\bauthor{\bsnm{C{\'o}rcoles}, \binits{A.D.}},
\bauthor{\bsnm{Temme}, \binits{K.}},
\bauthor{\bsnm{Harrow}, \binits{A.W.}},
\bauthor{\bsnm{Kandala}, \binits{A.}},
\bauthor{\bsnm{Chow}, \binits{J.M.}},
\bauthor{\bsnm{Gambetta}, \binits{J.M.}}:
\batitle{Supervised learning with quantum-enhanced feature spaces}.
\bjtitle{Nature}
\bvolume{567}(\bissue{7747}),
\bfpage{209}--\blpage{212}
(\byear{2019})
\end{barticle}
\endbibitem

\bibitem[\protect\citeauthoryear{Caro et~al.}{2022}]{caro2022generalization}
\begin{barticle}
\bauthor{\bsnm{Caro}, \binits{M.C.}},
\bauthor{\bsnm{Huang}, \binits{H.-Y.}},
\bauthor{\bsnm{Cerezo}, \binits{M.}},
\bauthor{\bsnm{Sharma}, \binits{K.}},
\bauthor{\bsnm{Sornborger}, \binits{A.}},
\bauthor{\bsnm{Cincio}, \binits{L.}},
\bauthor{\bsnm{Coles}, \binits{P.J.}}:
\batitle{Generalization in quantum machine learning from few training data}.
\bjtitle{Nature communications}
\bvolume{13}(\bissue{1}),
\bfpage{4919}
(\byear{2022})
\end{barticle}
\endbibitem

\bibitem[\protect\citeauthoryear{Huang et~al.}{2021}]{huang2021variational}
\begin{barticle}
\bauthor{\bsnm{Huang}, \binits{R.}},
\bauthor{\bsnm{Tan}, \binits{X.}},
\bauthor{\bsnm{Xu}, \binits{Q.}}:
\batitle{Variational quantum tensor networks classifiers}.
\bjtitle{Neurocomputing}
\bvolume{452},
\bfpage{89}--\blpage{98}
(\byear{2021})
\end{barticle}
\endbibitem

\bibitem[\protect\citeauthoryear{Park et~al.}{2020}]{park2020theory}
\begin{barticle}
\bauthor{\bsnm{Park}, \binits{D.K.}},
\bauthor{\bsnm{Blank}, \binits{C.}},
\bauthor{\bsnm{Petruccione}, \binits{F.}}:
\batitle{The theory of the quantum kernel-based binary classifier}.
\bjtitle{Physics Letters A}
\bvolume{384}(\bissue{21}),
\bfpage{126422}
(\byear{2020})
\end{barticle}
\endbibitem

\bibitem[\protect\citeauthoryear{Blank et~al.}{2020}]{blank2020quantum}
\begin{barticle}
\bauthor{\bsnm{Blank}, \binits{C.}},
\bauthor{\bsnm{Park}, \binits{D.K.}},
\bauthor{\bsnm{Rhee}, \binits{J.-K.K.}},
\bauthor{\bsnm{Petruccione}, \binits{F.}}:
\batitle{Quantum classifier with tailored quantum kernel}.
\bjtitle{npj Quantum Information}
\bvolume{6}(\bissue{1}),
\bfpage{41}
(\byear{2020})
\end{barticle}
\endbibitem

\bibitem[\protect\citeauthoryear{Schuld et~al.}{2020}]{schuld2020circuit}
\begin{barticle}
\bauthor{\bsnm{Schuld}, \binits{M.}},
\bauthor{\bsnm{Bocharov}, \binits{A.}},
\bauthor{\bsnm{Svore}, \binits{K.M.}},
\bauthor{\bsnm{Wiebe}, \binits{N.}}:
\batitle{Circuit-centric quantum classifiers}.
\bjtitle{Physical Review A}
\bvolume{101}(\bissue{3}),
\bfpage{032308}
(\byear{2020})
\end{barticle}
\endbibitem

\bibitem[\protect\citeauthoryear{Rethinasamy et~al.}{2023}]{rethinasamy2023estimating}
\begin{barticle}
\bauthor{\bsnm{Rethinasamy}, \binits{S.}},
\bauthor{\bsnm{Agarwal}, \binits{R.}},
\bauthor{\bsnm{Sharma}, \binits{K.}},
\bauthor{\bsnm{Wilde}, \binits{M.M.}}:
\batitle{Estimating distinguishability measures on quantum computers}.
\bjtitle{Physical Review A}
\bvolume{108}(\bissue{1}),
\bfpage{012409}
(\byear{2023})
\end{barticle}
\endbibitem

\bibitem[\protect\citeauthoryear{Bishop and Bishop}{2023}]{bishop2023deep}
\begin{botherref}
\oauthor{\bsnm{Bishop}, \binits{C.M.}},
\oauthor{\bsnm{Bishop}, \binits{H.}}:
Deep learning: Foundations and concepts
(2023)
\end{botherref}
\endbibitem

\bibitem[\protect\citeauthoryear{Scholkopf and Smola}{2018}]{scholkopf2018learning}
\begin{botherref}
\oauthor{\bsnm{Scholkopf}, \binits{B.}},
\oauthor{\bsnm{Smola}, \binits{A.J.}}:
Learning with kernels: support vector machines, regularization, optimization, and beyond
(2018)
\end{botherref}
\endbibitem

\bibitem[\protect\citeauthoryear{Bergholm et~al.}{2018}]{bergholm2018pennylane}
\begin{botherref}
\oauthor{\bsnm{Bergholm}, \binits{V.}},
\oauthor{\bsnm{Izaac}, \binits{J.}},
\oauthor{\bsnm{Schuld}, \binits{M.}},
\oauthor{\bsnm{Gogolin}, \binits{C.}},
\oauthor{\bsnm{Ahmed}, \binits{S.}},
\oauthor{\bsnm{Ajith}, \binits{V.}},
\oauthor{\bsnm{Alam}, \binits{M.S.}},
\oauthor{\bsnm{Alonso-Linaje}, \binits{G.}},
\oauthor{\bsnm{AkashNarayanan}, \binits{B.}},
\oauthor{\bsnm{Asadi}, \binits{A.}}, et al.:
Pennylane: Automatic differentiation of hybrid quantum-classical computations.
arXiv preprint arXiv:1811.04968
(2018)
\end{botherref}
\endbibitem

\bibitem[\protect\citeauthoryear{M{\"o}tt{\"o}nen et~al.}{2004}]{mottonen2004quantum}
\begin{barticle}
\bauthor{\bsnm{M{\"o}tt{\"o}nen}, \binits{M.}},
\bauthor{\bsnm{Vartiainen}, \binits{J.J.}},
\bauthor{\bsnm{Bergholm}, \binits{V.}},
\bauthor{\bsnm{Salomaa}, \binits{M.M.}}:
\batitle{Quantum circuits for general multiqubit gates}.
\bjtitle{Physical Review Letters}
\bvolume{93}(\bissue{13}),
\bfpage{130502}
(\byear{2004})
\end{barticle}
\endbibitem

\bibitem[\protect\citeauthoryear{Araujo et~al.}{2021}]{araujo2021divide}
\begin{barticle}
\bauthor{\bsnm{Araujo}, \binits{I.F.}},
\bauthor{\bsnm{Park}, \binits{D.K.}},
\bauthor{\bsnm{Petruccione}, \binits{F.}},
\bauthor{\bsnm{Silva}, \binits{A.J.}}:
\batitle{A divide-and-conquer algorithm for quantum state preparation}.
\bjtitle{Scientific reports}
\bvolume{11}(\bissue{1}),
\bfpage{6329}
(\byear{2021})
\end{barticle}
\endbibitem

\bibitem[\protect\citeauthoryear{Tacchino et~al.}{2019}]{tacchino2019artificial}
\begin{barticle}
\bauthor{\bsnm{Tacchino}, \binits{F.}},
\bauthor{\bsnm{Macchiavello}, \binits{C.}},
\bauthor{\bsnm{Gerace}, \binits{D.}},
\bauthor{\bsnm{Bajoni}, \binits{D.}}:
\batitle{An artificial neuron implemented on an actual quantum processor}.
\bjtitle{npj Quantum Information}
\bvolume{5}(\bissue{1}),
\bfpage{26}
(\byear{2019})
\end{barticle}
\endbibitem

\bibitem[\protect\citeauthoryear{Dunjko et~al.}{2016}]{dunjko2016quantum}
\begin{barticle}
\bauthor{\bsnm{Dunjko}, \binits{V.}},
\bauthor{\bsnm{Taylor}, \binits{J.M.}},
\bauthor{\bsnm{Briegel}, \binits{H.J.}}:
\batitle{Quantum-enhanced machine learning}.
\bjtitle{Physical Review Letters}
\bvolume{117}(\bissue{13}),
\bfpage{130501}
(\byear{2016})
\end{barticle}
\endbibitem

\bibitem[\protect\citeauthoryear{Tacchino et~al.}{2020}]{tacchino2020quantum}
\begin{barticle}
\bauthor{\bsnm{Tacchino}, \binits{F.}},
\bauthor{\bsnm{Barkoutsos}, \binits{P.}},
\bauthor{\bsnm{Macchiavello}, \binits{C.}},
\bauthor{\bsnm{Tavernelli}, \binits{I.}},
\bauthor{\bsnm{Gerace}, \binits{D.}},
\bauthor{\bsnm{Bajoni}, \binits{D.}}:
\batitle{Quantum implementation of an artificial feed-forward neural network}.
\bjtitle{Quantum Science and Technology}
\bvolume{5}(\bissue{4}),
\bfpage{044010}
(\byear{2020})
\end{barticle}
\endbibitem

\end{thebibliography}

\end{document}